\documentclass[a4paper]{article}
\usepackage{Odyssey2022}
\usepackage{epsfig,amssymb,graphicx,amsmath,booktabs,siunitx,xcolor,ifsym,pifont,amssymb,multirow,hyperref,cleveref}
\ninept

\newcommand{\snow}{\textcolor{blue}{\ding{100}}}%
\newcommand{\scratch}{\textcolor{gray}{\ding{48}}}%
\newcommand{\Fire}{\textcolor{red}{\textifsymbol{12}}}%

\title{An empirical study of weakly supervised audio tagging embeddings for general audio representations}

\name{Heinrich Dinkel,
    Zhiyong Yan,
    Yongqing Wang,
    Junbo Zhang,
      Yujun Wang}

\address{Xiaomi Corporation, Beijing, China
\\\{dinkelheinrich,yanzhiyong, wangyongqing3, zhangjunbo1,wangyujun\}@xiaomi.com}
\begin{document}
\maketitle

\begin{abstract}
We study the usability of pre-trained weakly supervised audio tagging (AT) models as feature extractors for general audio representations.
We mainly analyze the feasibility of transferring those embeddings to other tasks within the speech and sound domains.
Specifically, we benchmark weakly supervised pre-trained models (MobileNetV2 and EfficientNet-B0) against modern self-supervised learning methods (BYOL-A) as feature extractors.
Fourteen downstream tasks are used for evaluation ranging from music instrument classification to language classification.
Our results indicate that AT pre-trained models are an excellent transfer learning choice for music, event, and emotion recognition tasks.
Further, finetuning AT models can also benefit speech-related tasks such as keyword spotting and intent classification.
\end{abstract}

\section{Introduction}
The recognition of audio patterns is a vital part of enabling machines an understanding of our world.
Automatic audio pattern recognition is a broad research field, which can be split into two main categories: sound-related and speech-related.
Speech-related tasks mainly focus on the content of spoken languages i.e., who/what/when is something spoken? 
On the other hand, the sound domain focuses on classification, i.e., Bird songs, acoustic scenes, emotion, sound events.
Most prior works focused on either of those domains~\cite{Niizumi_BYOL,yang21c_interspeech,Kong2020d} i.e., models trained in the sound domain were also evaluated in the sound domain.

This work aims to provide an empirical study on the performance of models trained in the sound domain and evaluated on both sound and speech domains.
We believe that audio tagging (AT) is a reasonable pre-training task for sound and speech-domain tasks since AT incorporates unconstrained speech data, spoken in different languages and a variety of accents as well as sound events such as music.
Recent challenges such as the holistic evaluation of audio representations (HEAR) 2021~\footnote{For further information please visit \url{https://neuralaudio.ai/}.}~\cite{turian2022hear} have aimed to further push the efforts in researching universal audio embeddings.
% Our effort in this work focuses towards a similar goal to the HEAR 2021 challenge: Investigating the limits of a subset
To the best of our knowledge, there has been no comprehensive work that compared self-supervised approaches to their weakly supervised counterparts in the sound domain.
% More recently, the HEAR 2021 challenge~\cite{} has brought some insights into the potential 
Thus, this work aims to investigate the usage of AT pre-trained neural networks as a general audio representation, especially we aim to answer whether AT pre-trained models are useful for speech-related tasks.
Our contributions are as follows:
\begin{itemize}
    \item Explore AT pre-trained models as a common back-bone for sound and speech downstream tasks.
    \item A comparison of weakly supervised pretraining against modern self-supervised approaches.
    \item We provide insights into the limits of sound domain pretrained models to their application in the speech domain.
\end{itemize}

\section{Related work}
\begin{figure*}[tb]
    \centering
    \includegraphics[width=0.70\linewidth]{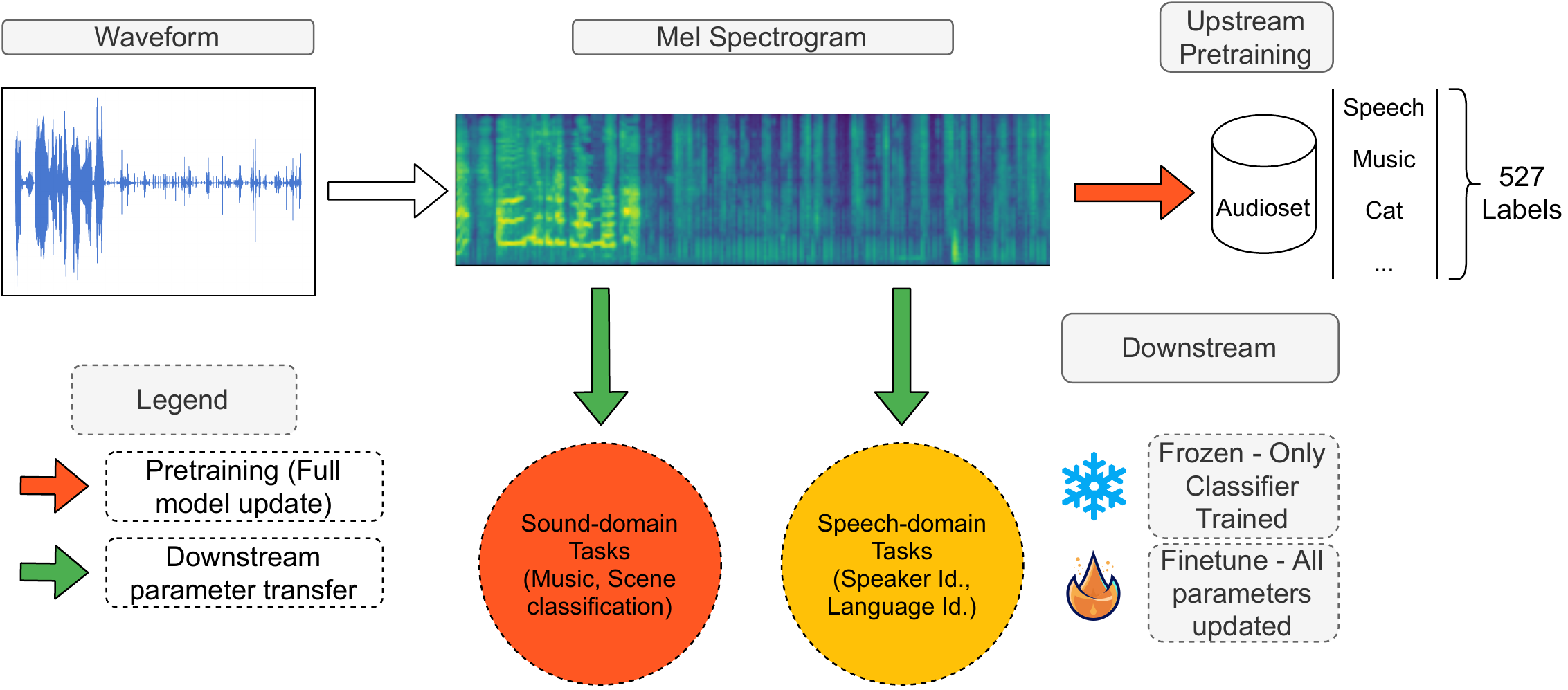}
    \caption{In this work, we investigate the use of (upstream) Audio tagging pre-trained models as general audio representations in two domains: sound and speech.}
    \label{fig:framework}
\end{figure*}

In terms of weakly supervised pre-training, the works in~\cite{Kong2020d,gong21b_interspeech,gong2021psla,htsat-ke2022} proposed novel model architectures for weakly supervised AT and explored the performance on a limited amount of downstream tasks.

Further, the works~\cite{saeed2021contrastive,wang2021multi,wang2021_multimodal,COALA_Xavier,tagliasacchi2020pre,gong2021ssast} explored self-supervised learning with a linear classifier for sound classification, focusing on sound-domain downstream tasks.
Initial work for cross-domain evaluation has been done in~\cite{Niizumi_BYOL,wang2021multi}, but those works mainly compared against other self-supervised or traditional weakly-supervised methods.
The work in \cite{kumar21_interspeech} explored the effects of weakly-supervised audio tagging pretraining to sound-domain downstream tasks, such as music tagging and acoustic scene classification.

The paper is organized as follows. 
\Cref{sec:method} describes our proposed framework and experimental details are laid out in \Cref{sec:experiments}.
Results are presented in \Cref{sec:results} and a conclusion is provided in \Cref{sec:conclusion}.

\section{Framework}
\label{sec:method}

Our approach trains a standard multi-label multi-class convolutional neural network (CNN) model on the \SI{5200}{h} long Audioset~\cite{gemmeke2017audio} containing 527 labels.
Given a training audio clip signal $\boldsymbol{x}$ (i.e., of length 10 s) and a multi-hot label $\boldsymbol{y} \in \{0,1\}^{527}$, the training objective is to predict $\boldsymbol{y}$ using a model $\mathcal{F}$, such that $\mathcal{F}(\boldsymbol{x}) \mapsto \boldsymbol{y}$. 
The model is trained to optimize the binary cross entropy (BCE), defined as:
\begin{equation}
    \mathcal{L}_{\text{BCE}} = \boldsymbol{y} \log \boldsymbol{\hat{y}} + (\boldsymbol{1} - \boldsymbol{y}) \log(\boldsymbol{1} - \boldsymbol{\hat{y}}),
\end{equation}
between the model predictions $\boldsymbol{\hat{y}}$, which are commonly processed using a sigmoid activation function, and the ground truth $\boldsymbol{y}$.
We use MobilenetV2~\cite{sandler2018mobilenetv2} (MBv2) and an Efficientnet-B0~\cite{tan2019efficientnet} (Eff-B0) as our neural network back-bones ($\mathcal{F}$), both of which are off the shelf models deployed in the vision domain and can be adapted to the sound domain with little effort.

Different from off the shelf available vision-based model, our audio-domain MBv2 and Eff-B0 networks use a decision-level output pooling function, which produces a single output vector $\hat{\boldsymbol{y}} \in \mathbb{R}^{527}$, from its per-timestep outputs $\hat{\boldsymbol{y}}_{1:T}$ as $\hat{\boldsymbol{y}} = \frac{1}{T}\sum_{t}^T\boldsymbol{\hat{y}}_t$ for the (pooled) audio duration $T$. 

Then, we explore transfer learning to downstream tasks by either freezing (only train a linear classifier) or finetuning (train entire model) the pre-trained CNN model, as can be seen in \Cref{fig:framework}.

\section{Experiments}
\label{sec:experiments}

\subsection{Datasets}

This work uses a single dataset for pretraining purposes, being Audioset~\cite{gemmeke2017audio} and 14 different downstream datasets for finetuning.
A short introduction for each dataset is provided.

\begin{table}[ht]
    \centering
    \begin{tabular}{l|rrrr}
        \toprule
        Dataset &  \#Train & \#Eval & \# Labels & Metric  \\
        \midrule
        Audioset~\cite{gemmeke2017audio} &  1,904,746 & 18,229 & 527 & mAP \\
        \hline
        US8k$^{\dag}$~\cite{salamon2014dataset} & 8,732  & -  & 10 & Accuracy \\
        FSD18~\cite{fonseca2017freesound} & 8,530 & 1,600 & 41 & mAP@3 \\
        FSD19-C~\cite{Fonseca2019audio} & 4,473 & 4,481 & 80 & \textit{lwl}wrap \\
        FSD19-N~\cite{Fonseca2019audio} & 17,833 & 4,481 & 80 & \textit{lwl}wrap \\
        FSD50k~\cite{fonseca2020fsd50k} & 36,796 & 10,231 & 200 & mAP \\
        ESC-50$^{\dag}$~\cite{piczak2015esc} & 2,000 & - & 10 & Accuracy\\
        NS~\cite{engel2017neural} & 283,704 & 4,097 & 10 & Accuracy \\
        MSoS~\cite{kroos2019generalisation} &  1,350 & 50 & 5 & Accuracy\\
        RAV$^{\dag}$~\cite{livingstone2012ravdess} & 1,400  & - & 8 & Accuracy\\
        \hline
        SCV1/11~\cite{warden2018speech} & 51,088 & 6,835 & 11 & Accuracy\\
        SCV1/30~\cite{warden2018speech} & 51,088 & 6,835 & 30 & Accuracy\\
        SCV2~\cite{warden2018speech} & 84,843 & 11,005 & 35 & Accuracy\\
        FSC~\cite{lugosch2019speech} & 23,132 & 3,793 & 31 & Accuracy\\
        Vox1~\cite{nagrani2017voxceleb} & 138,361 & 8,251 & 1251 & Accuracy\\
        VL~\cite{valk2021voxlingua107} & 2,536,954 & 1,609 & 107 & ErrorRate\\
         \bottomrule
    \end{tabular}
    \caption{Description of datasets used in this work with their respective evaluation metrics and their respective number of samples for training (\#Train) and evaluation (\#Eval). Datasets denoted with a $^{\dag}$ are k-fold cross-validated. Audioset is used as the upstream dataset for pretraining purposes exclusively. mAP represents mean average precision.}
    \label{tab:dataset}
\end{table}

\subsubsection{Sound domain datasets}

Three Freesound (FSD) datasets are used in this work, all of which focus on (multi-label) audio tagging.
We deploy FSDKaggle2018~\cite{fonseca2017freesound} (FSD18), FSDKaggle2019~\cite{Fonseca2019audio} (FSD19) and FSD50k~\cite{fonseca2020fsd50k} as our downstream datasets.
All recordings in the FSD datasets were obtained from Freesound and annotated with different vocabulary sizes.
FSD19 dataset has two optional training sets, a clean curated one (FSD19-C) and a large noisy one (FSD19-N).
ESC-50 is an environmental sound dataset~\cite{piczak2015esc} consisting of 50 distinct sound labels.
The Making Sense of Sounds~\cite{kroos2019generalisation} (MSoS) dataset focuses on classifying five categories types.
NSynth~\cite{engel2017neural} contains musical notes, labeled with the instrument family with overall 11 classes. 
The Ryerson Audio-Visual Database of Emotional Speech~\cite{livingstone2012ravdess} (RAVDESS) contains eight different emotions portrayed by professional actors.
UrbanSound8k~\cite{salamon2014dataset} (US8k) contains ten acoustic scene classes in urban environments.

\subsubsection{Speech domain datasets}

Keyword spotting (KWS) detects a fixed set of predefined keywords within a spoken utterance.
We use the Google Speech Commands (SPC) V1 and V2 datasets~\cite{warden2018speech} for this task, which contains 30 and 35 keywords, respectively.
For V1, we use the common 11 class subset, where the original 30 classes have been reduced to 10 common keywords +  ``Unknown''.
The largest downstream dataset in this work is the VoxLingua107~\cite{valk2021voxlingua107} (VL) dataset, containing over 6000 hours of data across 107 languages, which is aimed at language identification.
% Note that the evaluation portion consists of 33 verified languages.
Intent classification is evaluated on the Fluent Speech Commands~\cite{lugosch2019speech} (FSC) dataset, which requires categorizing utterances into (31) predefined slots.
Speaker identification (SID) is a classification task aiming to identify a person by their voice.
The VoxCeleb1~\cite{nagrani2017voxceleb} (Vox1) dataset is used for SID.
The data statistics and evaluation metrics for all downstream datasets are displayed in \Cref{tab:dataset}.

\subsection{Setup}

Regarding front-end feature extraction, all audio clips are converted to a 16 kHz sampling rate. 
As a common backbone, we use log-Mel spectrograms (LMS) with 64 bins extracted every 10 ms with a window of 32ms.
For all experiments, we apply batch-wise zero padding to the longest clip within a batch.
We use a batch-size of 32 for Audioset pretraining and 64 for all downstream tasks.
The neural network back-end is implemented in Pytorch~\cite{PaszkePytorch}.

\subsubsection{Audioset pretraining details}

For training our models on Audioset, a label-balancing sampling strategy is used~\cite{Dinkel2021TowardsDR,Kong2020d}, which selects samples from the training dataset such that each batch contains at least one label instance.
We use Adam optimization with a starting learning rate of 1e-4 using standard binary cross-entropy (BCE) as the criterion and a batch size of 32.

Audioset contains two training sets, being the 60 h long balanced set and the 5,200 h long unbalanced set.
We train our models on the full (balanced + unbalanced) training set for at most 100 epochs.
We checkpoint every $\approx \frac{1}{6}$ epoch, where the model is evaluated on the balanced subset.
An early stop of 5 checkpoints is adopted and the four best ranking validation checkpoints are weight averaged and tested on the public evaluation set.

Our augmentation procedure is split into wave-domain and spectral-domain.
In the wave-domain, we use polarity inversion, gain, and randomly shifting with their default settings as defined in torch-audiomentations\footnote{\url{https://github.com/asteroid-team/torch-audiomentations}}.
In the spectral-domain, we use mixup~\cite{zhang2018mixup} with $\beta(1, 1)$ and SpecAug~\cite{Park2019} with a time mask of 96 frames and a frequency mask of 12 bins.

\subsubsection{Downstream training details}

For all downstream tasks, we run the experiments for at most 150 epochs with an early stop of 15 epochs. 
Each downstream task saves the best model checkpoint given its respective main metric (see \Cref{tab:dataset}).
We use Adam optimization with a starting learning rate of 1e-4 with no learning rate scheduling.
After training, we weight-average the four best checkpoints on the held-out validation dataset and evaluate the averaged model on the evaluation set.
Note that for datasets that do not provide an independent cross-validation dataset (VL, MSoS), we use a 90\%/10\% split of the available development data for training and validation, respectively.
Further, for tasks that are evaluated over a K-Fold split (US8k, ESC-50, RAV), we use the pre-defined train/test split for training and cross-validation respectively.
On the K-Fold datasets, an average result across all folds is reported.

For tasks with variable-length input (FSD-50k, Vox1, VL, FSD2019), we preprocess the training dataset by first segmenting each training clip to trim to a maximal length.
We chose 10 seconds for FSD18 and FSDK50k datasets since their sample duration ranges between 0.3 and 30s and 4 seconds for the Vox1, FSD19, FSC, and VL datasets.
Note that each segment inherits its corresponding clip label.
We use a batch size of 1 during evaluation, in order to avoid any performance bias caused by zero-padding or clip-segmentation.
Note data augmentation is not applied for downstream tasks.
Cross-entropy is the default training criterion for all classification, which is replaced by BCE for multi-label datasets (FSD50k, FSD18, FSD19).

\section{Results}
\label{sec:results}

\begin{table}[ht]
    \centering
    \small
    \begin{tabular}{l|r|rrr}
    \toprule
    Model & \#Params (M) & mAP $\uparrow$ & mAUC $\uparrow$ & d' $\uparrow$  \\
        \midrule
    MBv2 & \textbf{3.0}& 40.6 & 97.1 & 2.695 \\
        Eff-B0 & 4.6 & 41.7 & 96.4 & 2.546 \\
        \hline
        \hline
        MBv1~\cite{Kong2020d} & 4.7  & 38.9 & 97.0 & 2.653 \\
        MBv2~\cite{Kong2020d} & 4.0 & 38.3 & 96.8 &  2.624 \\
        Eff-B2~\cite{gong2021psla} & 13.6 & 43.0 & - & - \\
        % AST~\cite{} & 80.2 & 36.6 & - & - \\
        % ResNet38~\cite{Kong2020d} & 73.7 & \textbf{43.4} & \textbf{97.4} & \textbf{2.737}  \\
        CNN14~\cite{Kong2020d} & 80.7 & \textbf{43.1} & \textbf{97.3} & \textbf{2.732}  \\
         \bottomrule
    \end{tabular}
    \caption{Baseline results of our models (rows 1-2) on the public Audioset evaluation set against other approaches (rows 3-6) from literature. Higher is better for all metrics and the best results are in boldface.}
    \label{tab:Audioset_ce_vs_bce}
\end{table}

\begin{table}[htbp]
    \centering
    \footnotesize
    \begin{tabular}{l|ccc||ccc}
        \toprule
        \multirow{2}{*}{Dataset} & \multicolumn{3}{c||}{MBv2} & \multicolumn{3}{c}{Eff-B0} \\
                                 & \scratch & \snow & \Fire & \scratch & \snow & \Fire \\
        \midrule
        MSoS $\uparrow$ & 56.85 & 90.40 & 93.55 & 58.80 & 89.80 & \textbf{93.75} \\
        NS $\uparrow$ & 33.56 & 66.50 & \textbf{79.93} & 78.82 & 64.65 & 77.92  \\
        US8K $\uparrow$ & 74.77 & 83.27 & 84.29 & 75.03 & 83.00 & \textbf{85.22}  \\
        RAV $\uparrow$ & 46.54 & 50.90 & 66.66 & 11.66 & 48.68 & \textbf{70.00} \\
        ESC-50 $\uparrow$ & 59.90 & 92.20 & 93.70 & 58.17 & 92.10 & \textbf{94.48} \\
        FSD18 $\uparrow$ & 85.60 & 87.31 & \textbf{94.43} & 78.96 & 91.02 & 91.12 \\
        FSD19-C $\uparrow$ &  56.16 & 68.84 & \textbf{72.96} & 54.95 & 67.86 & 72.92 \\
        FSD19-N $\uparrow$ & 38.17 & 53.57 & 49.97 & 38.46 &  \textbf{54.89} & 49.93 \\
        FSD50k $\uparrow$ &  38.03 & 44.41 &54.62 & 38.24  &44.66  & \textbf{57.20}\\
        \hline
        Vox1 $\uparrow$& 44.86 & 19.02 & 48.43 & \textbf{62.23} & 13.54 & 54.71 \\
        VL $\downarrow$&  {19.83} & 92.85 & 26.17 & \textbf{19.71} & 93.41 & 20.39 \\
        FSC $\uparrow$ & 99.12 & 17.37 & \textbf{99.47} & 97.94 & 13.58 & 98.04 \\
        SPCV1 $\uparrow$ & 97.62 & 69.28 & 98.27 & 97.75 & 68.76 & \textbf{98.30} \\
        % SPCV1/30 $\uparrow$ & 96.22 & 52.17 & 96.87 & 96.29 & 46.42 & \textbf{97.01} \\
        SPCV2 $\uparrow$ & 96.18 & 49.13 & 96.81 & 96.35 & 42.85 & \textbf{96.86}\\
        \bottomrule
 
    \end{tabular}
    \caption{Transfer learning results on the proposed downstream tasks. 
    Both models are either trained from scratch ({\scratch}), frozen ({\snow}), where only the classifier is trained, or finetuned ({\Fire}). 
    All results represent an average of 4 trials for each task (except VL). 
    The best result for each dataset is marked in bold. $\uparrow$ represents higher is better and $\downarrow$ lower is better.}
    \label{tab:transfer_learning}
\end{table}

\begin{table*}[!htbp]
    \centering
    \resizebox{0.99\textwidth}{!}{
    \begin{tabular}{m{1cm}l|rrrrrrrrrr}
    \toprule
    Paradigm & Method &  MSoS $\uparrow$ & NS $\uparrow$ & US8K $\uparrow$ & RAV $\uparrow$ & ESC-50 $\uparrow$ &  FSD18 $\uparrow$ & FSD19-C $\uparrow$ & FSD19-N $\uparrow$ & FSD50k $\uparrow$  \\
        \midrule
    \multirow{4}{*}{SSL} & BYOL-A~\cite{Niizumi_BYOL} & - & 74.10 & 79.10 & - &  - & - & - & - & - & \\
     & BYOL-A$^{*}$ \snow  & 78.90 & 75.27 & 79.78 & 65.48 & 86.45 & 90.06 & 64.07 & 36.28 & 31.10 \\
     & BYOL-A$^{*}$ \Fire  & 80.75 & \textbf{80.04} & 78.90 & 65.17 & 82.86 & 89.36 & 63.12 &  38.51 & 40.38 \\
      & COLA~\cite{Niizumi_BYOL} & - & 70.20 & 78.50 & - &  - & - & - & - & - & \\
    %   & CLAR~\cite{al2021clar} & - & 48.80 & - & - & 40.40 & - & -  \\
        \hline
    \multirow{7}{*}{Weak}   & WEANET~\cite{kumar2020sequential} & - & - & - & - & 94.10 & - & 72.80 & 50.30 & -  \\
    & CNN14~\cite{Kong2020d} \snow & 88.60 & - & - & 39.70 & 90.80  & - & - &- & -  \\        
    & CNN14~\cite{Kong2020d} \Fire & \textbf{96.00} & - & - & \textbf{72.10} & \textbf{94.70} & - & - & - & - &     \\
    % Weak  & AST~\cite{gong_interspeech} \Fire & - & - & - & - & 95.60  & - & \\
    \cline{2-12}
    & MBv2 \snow & 90.40 & 66.50 & 83.27 & 50.90 & 92.20 & 87.31 & 68.84 & 53.57 & 44.41  \\
      & MBv2 \Fire & 93.55 & 79.93 & 84.29 & 66.66 & 93.70 & \textbf{94.43} & \textbf{72.96} & 49.97 & 54.62  \\
    \cline{2-12}
      & Eff-B0 \snow  & 89.90 & 64.65 & 83.00 & 48.68 & 92.10 & 91.02 & 67.86 & \textbf{54.89} & 44.66 \\
      & Eff-B0 \Fire  & 93.75 & 77.92 & \textbf{85.22} & 70.00 & 94.48 & 91.12 & 72.92 & 49.93 & \textbf{57.20} \\
    \bottomrule
    \end{tabular}
    }
    \caption{Comparison of the proposed approach against other works in literature on sound-domain based tasks. Models denoted with $^{*}$ have been finetuned from a publicly available model. Best results in bold.}
    \label{tab:comparison_audio}
\end{table*}

\subsection{Upstream results on Audioset}
\label{ssec:results_audioset}

We provide a performance comparison of our baseline models against current state-of-the-art (SOTA) methods in \Cref{tab:Audioset_ce_vs_bce}.
Standard evaluation on Audioset includes mean average precision (mAP) as its most significant metric, whereas mean area under the curve (mAUC) and d-prime (d') are secondary metrics.
Our two utilized models provide excellent performance, especially when considering the few parameters.

\label{ssec:comparison}

\subsection{Transfer learning results}
\label{ssec:transfer_results}

The results of the proposed models for transferring to other tasks can be seen in \Cref{tab:transfer_learning}.
Results are categorized by their respective domain type (top = audio, bottom = speech).

As it can be seen, for all sound-related tasks, training a simple classifier on top of our pre-trained model (\snow) enhances performance again training from scratch (\scratch).
Large performance gains are observed for music/sound classification (MSoS, NS), audio tagging (FSD, ESC), and acoustic scene (US8k) datasets.
Smaller, but considerable gains are also observed for emotion detection (RAV).

Unfortunately, regarding most speech-related tasks (Vox1, SPCV1/2), both of our pre-trained model's performance largely drops if weights are frozen (\snow).
This can largely be explained due to the differences in audio pattern recognition and speech recognition: 
The spectral diversity for audio pattern tasks is much richer than for speech-related tasks.
Further, even though the pretraining dataset, Audioset, contains speech, the corresponding labels are coarse and have no connection to the content of speech e.g., ``Speech'' and ``Conversation'' are similar concepts of human language, but lack detailed content information.

Freeing all weights during the finetune process (\Fire), performance on SPCV1/2 and FSC datasets improve against the baseline from scratch.
This suggests that audio tagging pre-trained models are capable of enhancing performance on speech tasks, although the performance gains are limited.
Results on the VL dataset indicate that training from scratch is superior to a pre-trained alternative, which is likely due to the dataset's sufficient training size (6000 h).
% This is likely due to the large size (6000 h) of the VL dataset, which provides sufficient data samples to train the model well.

\subsection{Comparison to other approaches}

% We provide a comparison of our weakly pre-trained audio embeddings against other approaches in the literature for sound-related tasks in \Cref{tab:comparison_audio}.
We compare our weakly-supervised approach against the state-of-the-art CNN14 model and the modern self-supervised bootstrap your own audio latent (Byol-A)~\cite{Niizumi_BYOL} method for sound-related downstream tasks in \Cref{tab:comparison_audio}. 
We choose the publicly available BYOL-A model (``64x96d2048'') pre-trained on Audioset with a comparable parameter size of 5M. 

As the results in \Cref{tab:comparison_audio} indicate, pretrained SSL embeddings perform well on most downstream sound-tasks, but are mostly outperformed by weakly supervised approaches.
For the FSD19-N and FSD50k datasets, we observe a significant performance difference (38.51 vs. 54.89 and 40.38 vs. 57.20) between BYOL-A and our weakly trained Eff-B0.
When comparing our models against CNN14, we observe that both of our models obtain a superior performance when weights are frozen (MSoS, RAV, ESC-50), which indicates that higher mAP performance on Audioset might not transfer to other tasks.
Even though CNN14's fine-tuning performance is superior to our models, its parameter size of 80 M (see \Cref{tab:Audioset_ce_vs_bce}) slows down the finetuning process drastically, leading to comparatively long training times.

\begin{table}[htbp]
    \centering
    \resizebox{0.48\textwidth}{!}{
    \begin{tabular}{m{1mm}l|rrrrr}
    \toprule
     & Method &  Vox1 $\uparrow$ & VL $\downarrow$ & SPCV1 $\uparrow$ & SPCV2 $\uparrow$ & FSC  $\uparrow$ \\
        \midrule
    \multirow{6}{*}{\rotatebox[origin=c]{90}{SSL}} & BYOL-A~\cite{Niizumi_BYOL} & 40.10 & - & - & 92.20 & - \\  
      & BYOL-A$^{*}$ \snow & 30.53 & 94.10 & 93.11 & 90.67 & 67.10 \\
      & BYOL-A$^{*}$ \Fire & 57.91 & 34.18 & 97.89 & 96.65 & 97.78 \\
      & COLA~\cite{saeed2021contrastive} & 30.40 & - & - & 76.70 & -   \\
      & W2V~\cite{yang21c_interspeech} & 56.56 &  -  &  - & - & 84.92  \\
      & W2V2.0~\cite{yang21c_interspeech} & \textbf{75.18} &  -  & - & - & 84.92  \\
        \hline
    % \parbox[t]{2mm}{\multirow{6}{*}{\rotatebox[origin=c]{90}{Weak}}}  & CNN14$^{*}$ \snow &   & & 62.44 & 9.01 & 10.07 \\
    %  & CNN14$^{*}$ \Fire &  &  & \textbf{98.70} & \textbf{97.77} & \textbf{99.56} \\
    % \cline{2-7}
    % \cline{2-7}
     \parbox[t]{2mm}{\multirow{4}{*}{\rotatebox[origin=c]{90}{Weak}}} & MBv2 \snow & 19.02 & 92.85 & 69.28 & 49.13 & 17.37 \\
    & MBv2 \Fire & 48.43 & 26.17 & 98.27 & 96.81 & 99.47 \\
    \cline{2-7}
    & Eff-B0 \snow  & 13.54 & 93.41 & 68.76 & 42.85 & 13.58 \\
    & Eff-B0 \Fire  & 54.71 & \textbf{20.39} & \textbf{98.30} & \textbf{96.86} & \textbf{98.04} \\
    \bottomrule
    \end{tabular}
    }
    \caption{Comparison of our weakly-supervised approach against Wav2Vec (W2V), COLA and BYOL-A in literature on speech-domain based tasks. Models denoted with $^{*}$ have been finetuned from a publicly available model.}
    \label{tab:comparison_speech}
    
\end{table}

Further, results in \Cref{tab:comparison_speech} indicate that except for Vox1, finetuned weakly supervised pre-trained models can outperform SSL approaches by a large margin on the SPCV1/2 and FSC tasks.
Moreover, we observe a large performance gap between Wav2Vec and all other approaches on the Vox1 dataset, which indicates that our weakly-supervised pretraining and BYOL-A do not sufficiently process inter-utterance information, such as speech identity.

\section{Conclusion}
\label{sec:conclusion}

This paper explored an empirical evaluation of pre-trained AT models as possible candidates for general audio representations.
A MobileNetV2 and an EfficientNet-B0 were used as back-bones for finetuning on 14 datasets split into two domains.
The results indicate that within both sound and speech domains, AT pre-training can improve performance.

We observe that freezing pretrained parameters for tasks within the sound domain generally improves performance, while on the contrary models trained from scratch are preferred over models with frozen parameters.
However, when finetuning on speech datasets, we observe significant gains against training from scratch and even other self-supervised methods.
Given the results of this work, we provide suggestions regarding transfer learning within speech and sound domains:
\begin{itemize}
    \item For sound-domain tasks, AT-pretraining largely benefits performance and should therefore be preferred.
    \item AT-pretraining can be used for downstream speech-domain tasks only when parameter finetuning is possible.
    \item Self-supervised pretraining (BYOL-A) offers a well-rounded performance across both speech and sound domains without finetuning. However, additional finetuning only offers marginal performance improvements for most sound-domain tasks.
\end{itemize}
This paper provides empirical evidence that AT pre-trained models can be used across domains for speech-related tasks and displays the performance gap between current self-supervised solutions and weakly-supervised ones.

\bibliographystyle{IEEEbib}
\bibliography{Odyssey2022_BibEntries}

% This could be also done as follows:
%
%\begin{thebibliography}{10}
%\bibitem[1]{aluisio2001learn}Sandra M. Alu\'{i}sio, Iris Barcelos, Jandir Sampaio, and Osvaldo
%N. Oliveira Jr, ``How to learn the many unwritten
%``rules of the game'' of the academic discourse: a hybrid
%approach based on critiques and cases to support scientific
%writing,'' in Proceedings of the IEEE International Conference
%on Advanced Learning Technologies, Madison, USA,
%August 2001, pp. 257–260.
%\bibitem[2]{swales1987writing} John Swales and Hazem Najjar, ``The writing of research
%article introductions,'' Written communication, vol. 4, no.
%2, pp. 175–191, 1987.
%\bibitem[3]{day2012write} Robert Day and Barbara Pastel, How to write and publish
%a scientific paper, Cambridge University Press, 2012.
%\bibitem[4]{teufel2000} Simone Teufel, Argumentative zoning: information extraction
%from scientific text, Ph.D. thesis, University of Edinburgh,
%2000.
%\bibitem[5]{berkenkotter1989social} Carol Berkenkotter, Thomas N. Huckin, and John Ackerman,
%``Social context and socially constructed texts: The
%initiation of a graduate student into a writing research community.
%technical report no. 33.,'' Tech. Rep., Center for
%the Study of Writing, University of California Berkeley \&
%Carnegie Mellon University, 1989.
%\end{thebibliography}

\end{document}